\documentclass[12pt,preprint,a4paper]{aastex}
\usepackage{graphics,epsf}
\usepackage{epstopdf}
\usepackage{multirow}
\usepackage{amsmath}                % American Mathematical Society package
\usepackage{amsfonts}               % American Mathematical Society fonts
\usepackage{amssymb}                % American Mathematical Society symbol
\usepackage{epsfig}                 % EPS figures
\usepackage{float}                  % Placing Figures
\usepackage{tabularx} 
\def \eV{~\rm{eV}}

\def \km{~\rm{km}}

\def \erg{~\rm{erg}}
\def \yrs{~\rm{yrs}}

\def \keV{~\rm{keV}}

\makeatletter
\newcommand*{\rom}[1]{\expandafter\@slowromancap\romannumeral #1@}
\makeatother

\date{\today}

\begin{document}

\title{The imprints of the last jets in core collapse supernovae}

\author{Ealeal Bear\altaffilmark{1}, Aldana Grichener\altaffilmark{1} and Noam Soker\altaffilmark{1}}

\altaffiltext{1}{Department of Physics, Technion -- Israel Institute of Technology, Haifa
32000, Israel; ealealbh@gmail.com; aldanag@campus.technion.ac.il; soker@physics.technion.ac.il}

\begin{abstract}
We analyze the morphologies of three core collapse supernova remnants (CCSNRs) and the energy of jets in other CCSNRs and  in Super Luminous Supernovae (SLSNe) of type Ib/Ic/IIb, and conclude that these properties are well explained by the last jets' episode as expected in the jet feedback explosion mechanism of core collapse supernovae (CCSNe).
The presence of two opposite protrusions, termed ears, and our comparison of the CCSNR morphologies with morphologies of planetary nebulae strengthen the claim that jets play a major role in the explosion mechanism of CCSNe. 
We crudely estimate the energy that was required to inflate the ears in two CCSNRs, and assume that the ears were inflated by jets. We find that the energies of the jets that inflated ears in 11 CCSNRs span a range that is similar to that of jets in some energetic CCSNe (SLSNe), and that this energy, only of the last jets' episode, is much less than the explosion energy. This finding is compatible with the jet feedback explosion mechanism of CCSNe, where only the last jets, that carry a small fraction of the total energy carried by earlier jets, are expected to influence the outer parts of the ejecta. We reiterate our call for a paradigm shift from neutrino-driven to jet-driven explosion models of CCSNe.  
\end{abstract}


% ==========================================================
\section{INTRODUCTION}
\label{sec:intro}
% =========================================================

In a rich variety of astrophysical objects a compact object accretes mass from an extended ambient gas and launches jets. The jets heat and/or expel the ambient gas, hence reducing the accretion rate, and by that set a negative feedback cycle. 
Similar properties of this jet feedback mechanism (JFM) between  different types of systems exist \citep{Soker2016Rev}. Core collapse supernova (CCSN) explosions are events where the JFM might play a decisive role.  
Support to the notion that jets play a role in the explosion of many CCSNe 
comes from observations (e.g., \citealt{Lopezetal2014, Gonzalezetal2014, Milisavljevicetal2015, Tanakaetal2017}) and  theoretical arguments, including numerical simulations (e.g. \citealt{BrombergTchekhovskoy2016, Gilkis2017, Nishimuraetal2017}). 

Most of the above listed theoretical studies assume that the pre-collapse core of the exploding star must be rapidly rotating for the newly born neutron star (NS; or black hole) to launch jets. 
Such a degree of rapid rotation can result only by a stellar binary interaction, most likely through a common envelope evolution. As such an evolutionary route is rare, these studies consider jet-driven explosions to be rare, and take place in specific type of CCSNe. These studies assume that the majority of CCSN explosions are driven by neutrinos (e.g., \citealt{Mulller2016}, for a recent review).  

We instead adopt the view that all CCSNe explode by jets which operate through the JFM (e.g., \citealt{Gilkisetal2016}). When the pre-collapse core angular velocity is low, pre-collapse turbulence \citep{GilkisSoker2016} and post-collapse instabilities \citep{Papishetal2015a} supply stochastic angular momentum to the gas accreted onto the newly born NS.
In that case the axis of the jets changes direction and the star explodes by jittering jets \citep{Soker2016Rev}.  
The post collapse instabilities are mainly the spiral modes of the standing accretion disk instability (SASI; \citealt{Papishetal2015a}), that have been shown to be very persistent and vigorous (e.g., \citealt{ Jankaetal2016, MorenoMendezCantiello2016, Blondinetal2017, Kazeronietal2017}). \cite{SchreierSoker2016} suggest that even if the temporarily specific angular momentum falls somewhat below the value required to form an accretion disk, jets might still be launched. 

Several studies have suggested a link between jets in energetic explosions and in weaker explosions (e.g., \citealt{Sobacchietal2017}). 
\cite{Marguttietal2014}, for example, find that their observations support theoretical proposals that link relativistic SNe with the weakest observed engine-driven SNe, where they argue that jets barely fail to break out. 

In light of these considerations, and the difficulties of the neutrino-driven mechanism (e.g., \citealt{Kushnir2015, Papishetal2015b, Janka2017}), our group (e.g., \citealt{Papishetal2015b, Soker2017a}) has called for a paradigm shift from neutrino-driven explosions to jet-driven explosions of all CCSNe; \cite{Piranetal2017} followed with a similar but a weaker call. 

In the present paper we analyze some observations under the assumption that all CCSN explosions are driven by jets. In section \ref{sec:last} we discuss the imprints of the last episode of jets to be launched, in section \ref{sec:morphology} we study three supernova remnants (SNRs) that we did not study before for last jets, and in section \ref{sec:energy} we compare the energy of the last jets in several CCSNRs and energetic CCSNe (SLSNe). 
We summarize our main results in section \ref{sec:summary}. 

% ==========================================================
\section{THE LAST EPISODE OF JETS}
\label{sec:last}
% ==========================================================
 
 According to the JFM (e.g., \citealt{Gilkisetal2016}) as long as mass is accreted from the collapsing core onto the newly born NS or black hole, jets are launched. After the jets remove the entire core, the accretion ceases and the jets are turned off. However, for a short time after the removal of the entire core some mass is still flowing inward near the NS. This mass launches the last jets in a last jet-launching episode. The last jets expand into an already exploding core, and therefore they are more free to expand and interact with the outer parts of the envelope. Namely, while during most of the jet-launching period the jets collide with the inner parts of the star (inside about $10^4 \km$; \citealt{PapishSoker2011}) and do not leave an imprint on the outer parts, the last jets might interact with the outer parts of the stars \citep{Gilkisetal2016}. This has the following implications. 

\textit{1. The formation of ears.} As we discussed in previous papers \citep{GrichenerSoker2017, BearSoker2017, Soker2017c}, the last episode of jets can inflate two ears. Ears are defined as two opposite protrusions from the main SNR shell. The last episode of jets might flow freely to the edge of the expanding envelope, and gently breakout, leaving the imprint of two opposite ears.  The jets leave the NS in opposite directions relative to the NS. However, the ears need not be exactly opposite and equal to each other. Firstly, if the NS has a natal kick (that is not along the axis of the jets) the two jets will have a tangential velocity component (relative to their axis) in the same direction. Secondly, the expanding stellar gas is expected to be clumpy because of instabilities in the explosion process, and because of the earlier jets. Hence, the two jets will encounter different environments. The outcomes of these processes are that the two opposite and identical jets at launching, do not necessarily form identical and exactly opposite ears. 
We note that the ears are formed by the last jets that do not encounter the dense core of the star. They are launched by the mass that is accreted on to the central NS after the core has been exploded. We here raise the possibility that in some cases this last accreted material leads to to two jets'-launching episodes, that in turn might form two pairs of ears that are inclined to each other.
We study two SNRs with ears in section \ref{sec:morphology}. 

 In relation to the last episode of jets, we note the followings. In a fallback flow, material that initially moves out later falls back onto the NS. In the last episode considered here the gas always falls in. Nonetheless, we do note that \cite{Utrobinetal2015} performed 3D explosion simulations (see also \citealt{Rantsiouetal2011}), and did not find that jets are formed by the accreted fallback gas. Our proposed scenario implies that the launching of jets is a complicated process that requires more ingredients in the numerical simulations, e.g., magnetic fields.  
 
\textit{2. Lower than explosion energy.} The last jets carry only a small fraction of the explosion energy (see section \ref{sec:energy}). Since only the last jets leave any imprint on the outer regions, any estimate of the jets' energy that is based on the imprints of the jets on the outer parts of the expanding gas will yield an energy much lower than the explosion energy. Therefore, the finding that the imprints of jets in the outer regions of the expanding gas result from energy lower than the explosion energy cannot be used as an argument against the jet-driven explosion, but as an expected result. 

\textit{3. Departure from spherical symmetry in the outer parts.} In the jittering jets explosion mechanism at each jets'-launching episodes the two jets are launched along the same symmetry axis but in opposite directions relative to the NS (or black hole). This symmetry axis, however, is different for different jets'-launching episodes \citep{PapishSoker2011}. Namely, the symmetry axis of the two opposite jets has different directions at different jet-launching episodes. 
The explosion has a global spherical morphology. The last jets are launched along one direction, and they have the largest imprint on the morphology of the outer regions. For that, in the jittering jet explosion mechanism the explosion might have a global spherical inner region, and a general bipolar outer region. We suggested that this general flow structure accounts for the morphology of the SNR Cassiopeia A \citep{Gilkisetal2016, Soker2017c}. Another CCSN that this flow structure might account for is SN~2014ad. 
From their spectropolarimetric data, \cite{Stevanceetal2017} conclude that SN~2014ad has a nearly spherical interior, and an aspherical outer ejecta. 
They, then argue that it is difficult to reconcile the geometry of the deeper ejecta with a jet-driven explosion. We, on the other hand, suggest that the jittering jets explosion mechanism might explain this morphology.

%==========================================================
\section{IMPRINTS OF LAST JETS IN THREE SNRs }
\label{sec:morphology}
%========================================================

In this section we analyze the structure of three CCSNRs: G292.0+1.8, N49B and RCW 103. The CCSNRs G292.0+1.8 and N49B contain two protrusions from the main SNR shell which sustain the required properties in order to be considered as ears, according to the criteria listed by \cite{GrichenerSoker2017}. We present images of these CCSNRs, on which we mark some morphological features that we will use in section \ref{sec:energy} to calculate the ratio of the energy that was required to inflate the ears to the total explosion energy. 

In both images we use a green double-headed arrow to mark the diameter of the base of the ear on the SNR, a red double-headed arrow to mark the diameter of the ear, and an orange double-headed arrow to mark the diameter of the SNR. For CCSNR RCW 103 which has no observed ears, we chose a different method. We compare its morphology to planetary nebulae (PNe) that share the same observed morphology features, and in addition have jets (e.g., \citealt{Tocknelletal2014} and a review by \citealt{Soker2016Rev}). We use these similarities to argue that the SNR RCW 103 was shaped by jets that are not active any more. 
 
%==========================
\subsection{G292.0+1.8}
%==========================

\begin{figure} %[H]
\centering
\hskip -6.00 cm
\includegraphics[width=22cm]{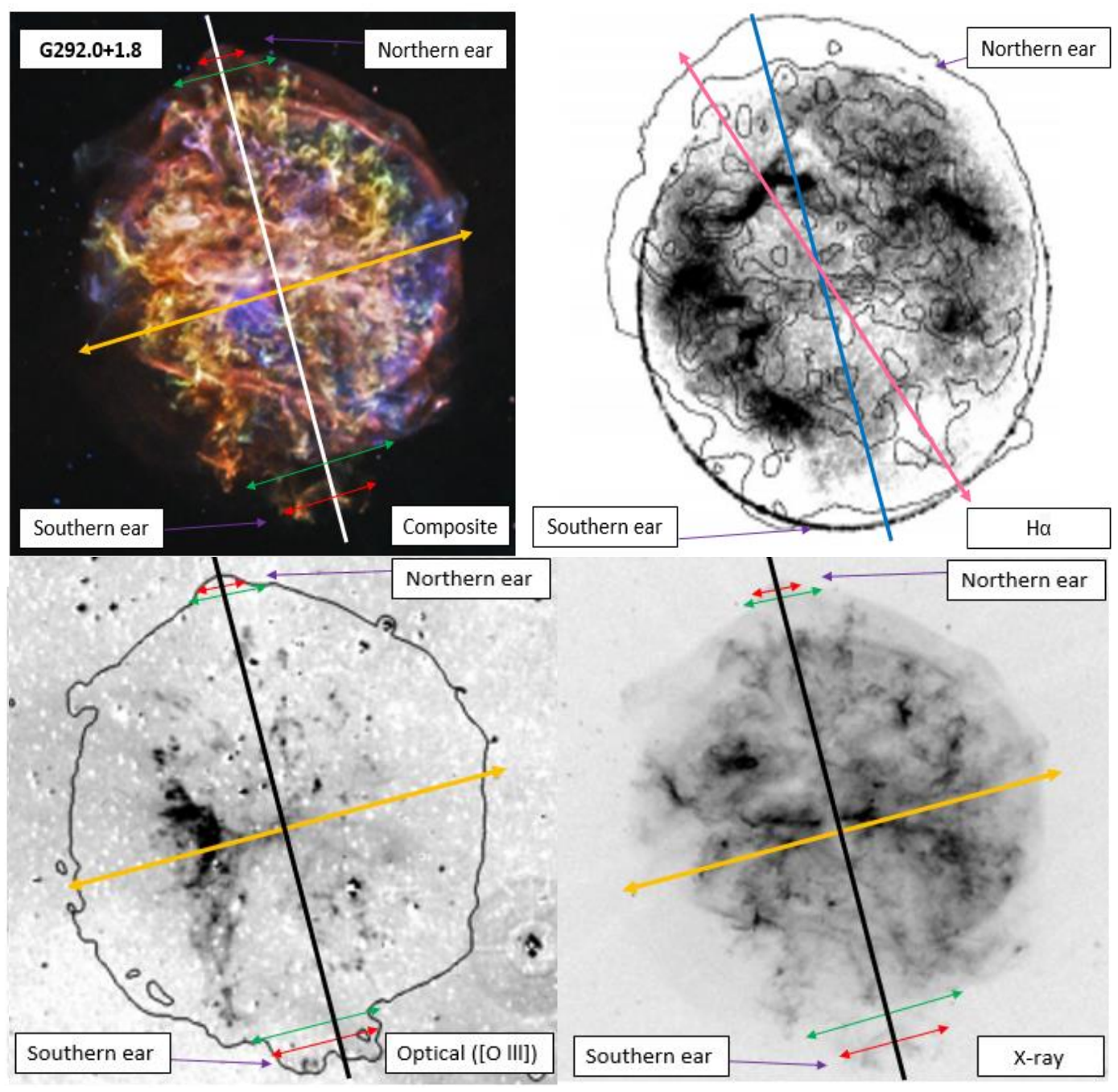}\\
\vskip -2.00 cm
\caption{Images of the CCSNR G292.0+1.8 in various wavelengths. In all panels the line from upper left to lower right is our assumed symmetry axis that connects the two ears.
    \textit{Upper left panel:} A composite image taken from the Chandra gallery (based on \citealt{Parketal2007}). Red, orange, green and blue colors represent O~Ly$\alpha$ combined with Ne~He$\alpha$ ($0.58 - 0.71$ and $0.88 - 0.95 \keV$), Ne~Ly$\alpha$ ($0.98 - 1.10 \keV$), Mg~He$\alpha$ ($1.28-1.43 \keV$), and Si~He$\alpha$ combined with S~He$\alpha$ ($1.81 - 2.05$ and $2.40 - 2.62 \keV$), respectively. White represents the optical band.   
    \textit{Upper right panel:} Zero velocity H$\alpha$ image taken from \cite{Ghavamian2005}. The blue solid line connects the two opposite ears, while the pink solid double-headed arrow is a possible symmetry axis defined by the two bright H$\alpha$ arcs. 
    \textit{Lower left panel:} An optical ([O~\rom{3}]) image taken from \cite{WinklerLong2006} and reproduced by \cite{Ghavamian2012}. 
    \textit{Lower right panel:} A 510 Ks Chandra X-ray image in $0.3-8.0 \keV$ taken from \cite{Parketal2007} and reproduced by \cite{Ghavamian2012}.} 
\label{G292.0+1.8}
\end{figure}

In Fig. \ref{G292.0+1.8} we present the structure of the CCSNR G292.0+1.8 in several wavelengths, and mark the relevant morphological features. 
Although small, we can note the two opposite ears on most panels, as marked. We assume that these ears were inflated by the last jets that are long gone. As in our earlier papers, we take the direction of the two opposite dead jets to be along the line connecting the two ears, as marked in all panels (by a white line in the upper left panel, by a blue line in the upper right panel, and by black lines in the lower panels).
%FFFFFFFFFFFFFFFFFFFFFFFF
%FFFFFFFFFFFFFFFFFFFFFFFF
 
 On the upper right panel of Fig. \ref{G292.0+1.8} we marked two possible symmetry axes. The blue line connects the two ears, as in the other three panels. The pink line marks the symmetry axis of the two bright H$\alpha$ arcs. This line is based on the morphology of many PNe that have a similar structure and in addition have ears and/or jets, as we discuss in section \ref{subsec:RCW103} below. As the two symmetry lines are very close to each other, we take it to imply that the same two opposite jets shaped the ears and the bright H$\alpha$ arcs. We further discuss this in relation to the PN NGC~40 in section \ref{subsec:RCW103}. 
 
%==========================
\subsection{N49B}
\label{subsec:N49B}
%==========================

\begin{figure} %[H]
\includegraphics[width=13cm]{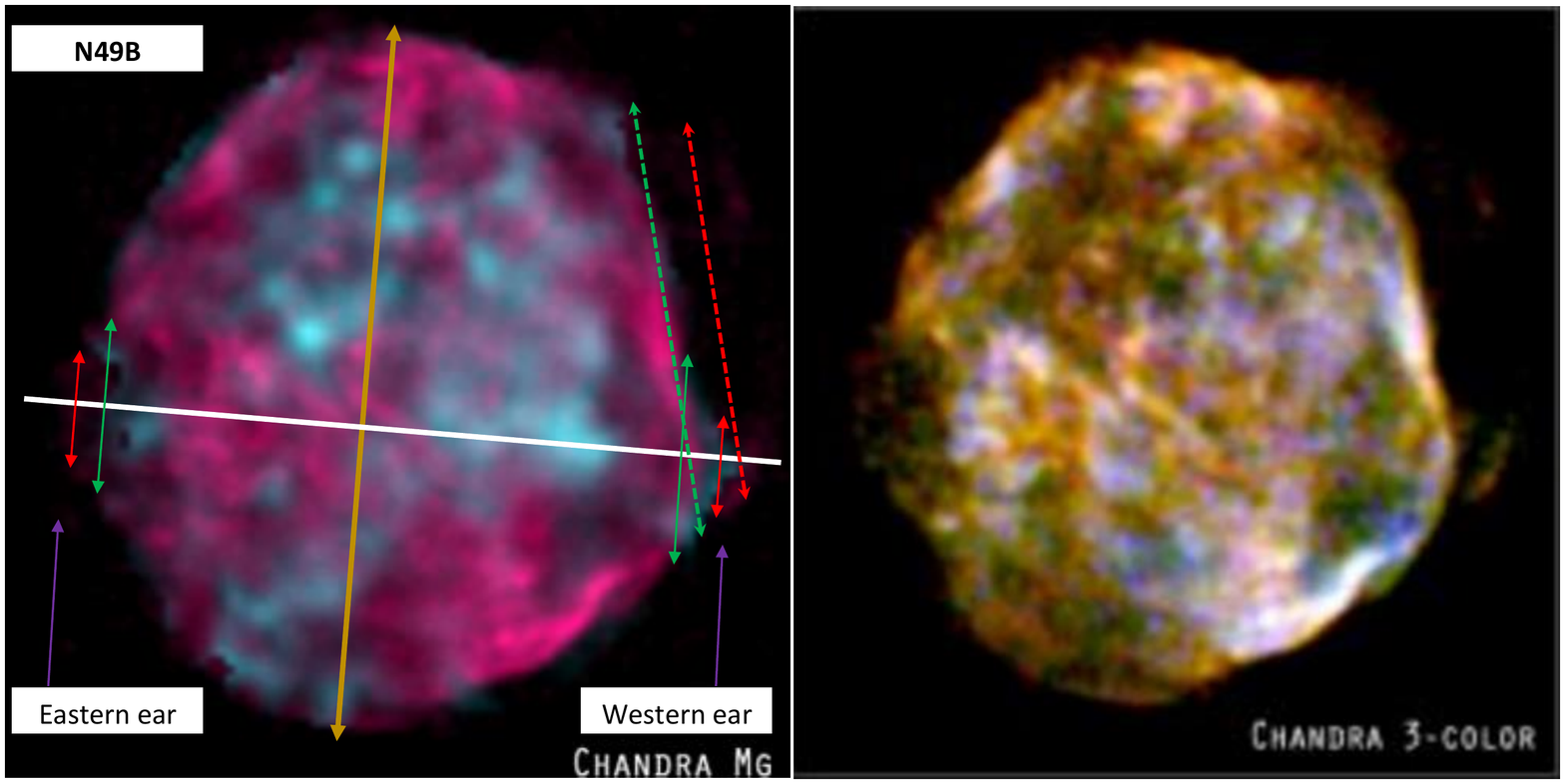}\\
%\vskip -6.20 cm
\centering
\caption{ The left panel shows a broadband image of N49B taken from the Chandra gallery (based on \citealt{Parketal2003}). The Pink color is composed of the energy bands $0.3-3.0 \keV$, while the cyan represents a Mg line ($1280 - 1440 \eV$). The white line marks the symmetry axis of the equal-size ears, which is perpendicular to the radius of the SNR (marked by the orange double-headed arrow). The solid green double-headed arrows mark the locations of the diameter of the base of each of the two visible and well defined opposite ears. The red double-headed arrows mark the diameter of those ears. The dashed double-headed arrow represents the corresponding quantities for a larger possible western ear. The right panel is an X-ray image of N49B with no marks. In this panel the red, green, and blue colors represent the energy bands $0.3-0.8 \keV$, $0.8-1.5 \keV$, and $1.5-3.0 \keV$, respectively. Another image showing the large western ear is given by \cite{Williamsetal2006}.    
} 
\label{N49B.eps}
\end{figure}

The ears of the SNR N49B are complicated. In the left panel of Fig. \ref{N49B.eps} we present two possibilities for the western ears. In the first case the ears are almost symmetric in size and are represented by the solid green and red double-headed arrows. In the second case the western ear is much larger and has a major offset to the north as marked by the dashed double-headed arrows. The eastern ear can either be the same as the one we see and is marked, or possibly a larger symmetrical eastern ear exists, but it is too faint to be observed. Other possibilities are that it has already been dispersed, or it is projected on the main SNR. The later explanation is quite possible, as it seems that the line that connects the two equal size ears, the white solid line, is tilted with respect to the plane of the sky, i.e., the eastern ear might be projected on to the much brighter main shell, hence undetectable.
%FFFFFFFFFFFFFFFFFFFFFFFF
%FFFFFFFFFFFFFFFFFFFFFFFF

One possible explanation to the presence of both large and small western ears is an interaction with the ISM. Another possibility is that the jets from the last two jet-launching episodes made their imprints on the SNR, and the two jets' symmetry axes were tilted with respect to each other.

%==========================
\subsection{RCW 103}
\label{subsec:RCW103}
%==========================
 
\begin{figure}  %[H]
\includegraphics[width=15cm]{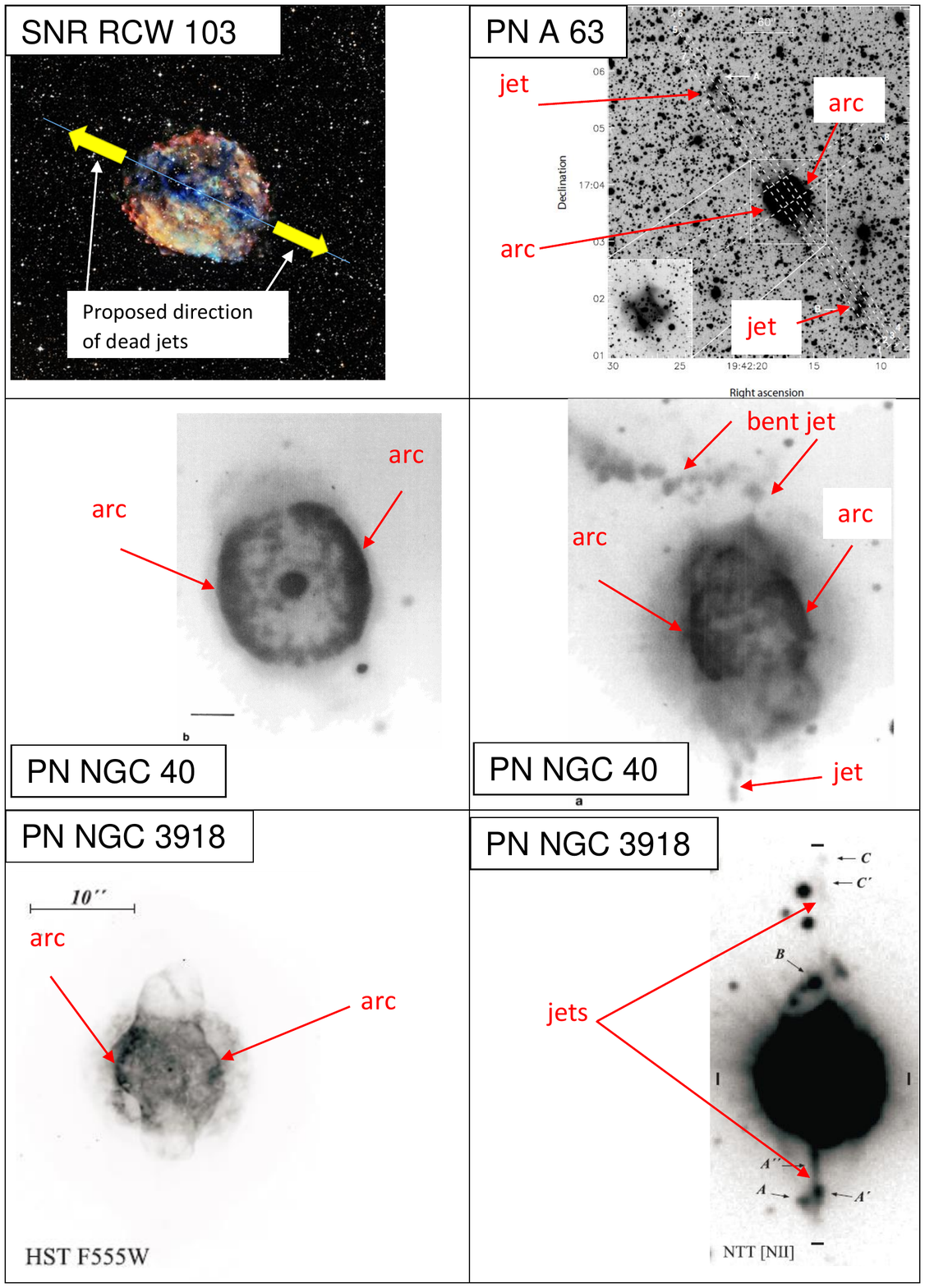}\\
\vskip -2.20 cm
\centering
\caption{A comparison of similar morphological features between the CCSNR RCW~103 and three PNe, A~63 (image taken from \citealt{Mitchelletal2007}), NGC~40 (both images are from \citealt{Meaburnetal1996}), and NGC~3918 (both images are from \citealt{Corradietal1999}; images are rotated so that the jets are vertical).  
The image of RCW~103 is a composite X-ray image in three energy bands (low=red, medium=green, highest=blue) combined with an optical image from the Digitized Sky Survey (image taken from the Chandra website based on \citealt{Reaetal2016}). 
The proposed original directions of the, already dead, jets in the CCSNR RCW~103 are marked by yellow thick arrows.} 
\label{RCW103}
\end{figure}

  In Fig. \ref{RCW103} we present the CCSNR RCW 103 along with three PNe. We see no ears in the CCSNR RCW 103 but present it here as we attribute some of its morphological features to the last jets in the frame of the jittering-jets explosion mechanism. Based on its morphological similarity to many PNe, we suggest that it was shaped by jets that inflated ears. Either the ears in RCW 103 are too faint to be observed, or they have already been dispersed in the ISM. 
In Fig. \ref{RCW103} we present three PNe that have two arcs and jets but no ears. All the PNe are thought to have been shaped by jets. We emphasize again the similarity between these PNe and RCW 103 (as well as other CCSNe), with respect to the role of the last jets. In PNe the main part of the nebula was ejected by the asymptotic giant branch stellar progenitor. In the PNe we present here, the jets have shaped some parts of this nebula. In CCSNe, the main mass of the ejecta was ejected during the explosion. We are only examining the role of the last jets in shaping some parts of the ejecta, and later the main SNR shell that contains also the swept-up ISM. The shaping of the main nebula/shell by the final jets is similar in the PNe we present here and the CCSNRs we study here, including RCW~103 that has no observable ears. 
%FFFFFFFFFFFFFFFFFFFFFFFF
%FFFFFFFFFFFFFFFFFFFFFFFF

The PNe presented in Fig. \ref{RCW103} and their relevant properties are as follows. 
\newline
\textit{Abell 63. } The structure of A~63 demonstrates the presence of two bright arcs and far-away two opposite faint jets. The jets shaped the inner bright part to have a hollow cylindrical shape, that on projection reveals two arcs. 
For more PNe that show jets together with a hollowed central bright nebula see for example \cite{GarciaDiazetal2012}, \cite{Clarketal2013} and \cite{Clyneetal2015}. 
Had the jets of A~63 been a little fainter, we would not have detected them. Presented is a deep H$\alpha$ + [N II]~λ6584\AA~image taken from \cite{Mitchelletal2007}, that is  based on \cite{PollaccoBell1997}.

\textit{NGC 40.} The left and right images of NGC~40 in Fig. \ref{RCW103} emphasize the arcs and jets, respectively. The jets are much fainter than the arcs, and in the future it is likely that the jets will not be detected any more, only the bright nebula with its arcs. Interestingly, the line connecting the southern jet with the northern one before it is bent, is somewhat tilted with respect to the symmetry axis of the arcs (a line through the center parallel to the arcs). Earlier jets could have had different directions, like precessing jets, and hence the arcs have a different symmetry axis. In any case, the jets shaped the arcs despite the tilted direction. 
The two images of NGC~40 are taken from \cite{Meaburnetal1996}, and both are H$\alpha$ + [NII] 6584\AA~ intensity maps.
The right panel of NGC~40 is in a logarithmic scale (see also  
\citealt{Balicketal1992} on this PN, and on NGC~6804 that also presents arcs and polar outflows).

The tilted symmetry axes in NGC~40 teach us also on the CCSNR~G292.0+1.8, where the symmetry axis of the ears is somewhat tilted with respect to the symmetry axis of the arcs (upper right panel of Fig. \ref{G292.0+1.8}). The PN NGC~40 strengthens our claim that jets shaped both ears and arcs in SNR~G292.0+1.8.

\textit{NGC~3918.} As with NGC~40, the left and right images emphasize the arcs and jets, respectively. The images are taken from \cite{Corradietal1999} who study the jets in NGC~3918, and where more details of this PN and its observations are given 
(see also \citealt{Corradietal2003}). For us the important structures are the arcs and jets that share a symmetry axis, and as with the other PNe, the jets are much fainter. The fainter jets imply that in many cases the jets are below detection limit, and only the arcs are observed. This is our claim for the CCSNR RCW~103.

We summarize this subsection by restating our claim, based on the similarities presented in Fig. \ref{RCW103} between the CCSNR RCW~103 and three PNe (and there are many more PNe with similar morphologies), that jets shaped the structure of RCW~103. However, unlike the case when ears are observed, in this case we have no information to calculate the energy of these jets. This conclusion is based on the assumption that the elongated structures along the symmetry axis of PNe were formed by jets (see review by \citealt{Soker2016Rev}).
 
%==========================================================
\section{THE ENERGY OF THE LAST JETS }
\label{sec:energy}
%========================================================

To calculate the energy that was required to inflate the ears relative to the kinetic energy of the SNR shell, we follow the procedure used by \cite{GrichenerSoker2017}, where more details can be found. We here briefly list only the basic assumptions of that procedure. 
(1) The SNR was spherical before jets shaped the ears. (2) The ejecta (that together with the swept-up ISM later forms the main shell) was uniform with a constant density per surface unit area. (3) The mass in the jets was small relative to the mass of the ejecta of the SNR, and their velocity higher than that of the ejecta. (4) The jets shaped the ears shortly after the explosion. (5) Each ear has a more or less hemi-spherical shape. For that we need the base and radius of each ear, as marked on Figs. \ref{G292.0+1.8} and \ref{N49B.eps} by double-headed arrows. (6) The velocities of the ears and SNR ejecta did not change since the formation of the ears. (7) The SNR expands into  a uniform ISM.  
We do note that not all assumptions hold for all CCSNRs. For example, the CCSNR W44 expands into a nonuniform medium. Also, in a recent paper \cite{ZhouVink2017} raise the possibility that W49B might be a remnant of a type Ia SN (if this claim holds, then W49B does not belong to our study). 

Under these simplifying assumptions, we calculate the extra energy that was required to accelerate the portion of the ejecta that was shaped by the jet to form an ear, $\Delta E_{\rm ear}$, relative to the initial kinetic energy of the entire ejecta, which under our assumption is the explosion energy $E_{\rm SNR}$ (technical details can be found in \citealt{GrichenerSoker2017}) 
\begin{equation}
\epsilon_{\rm ear} \equiv \frac{\Delta E_{\rm ear}}{E_{\rm SNR}} . 
\label{eq:epsilon1}
\end{equation}
As not all the kinetic energy of the jet was deposited in the ear that the jet inflated, i.e., some energy was radiated away, the energy of the jet was somewhat larger than the present energy in the ear $E_{\rm jet} \ga \Delta E_{\rm ear}$. 

In Table \ref{table:calculations} we list the value of $\epsilon_{\rm ear}$ for each of the two ears of the 10 SNR, and for one ear in the Crab Nebula. These values for the first 8 SNRs are taken from \cite{GrichenerSoker2017}, for W49B from \cite{BearSoker2017}, and the last two SNRs are the new ones studied here. For N49B we list the energy both for the western small ear and for the western large ear.
We note that one of these SNR is the Vela SNR which was recently suggested by \cite{Garciaetal2017} to contain a jet like structure similar to SNR Cass A.

The quantity $\epsilon_{\rm ears}$ given in the fourth column is the sum of the values for the two ears. In columns fifth and sixth we list the minimum, $E_{\rm SN, min}$, and maximum, $E_{\rm SN, max}$, explosion energies (in units of $10^{51} \erg$) that we found in the literature (listed in the last column). 
The seventh column gives the age of the CCSNR.  
% TTTTTTTTTTTTTTTTTTTTTTTTTTTTTTTTTTTTTTTTTTTTTTTTTTT
\begin{table}[bp]
%\begin{center}
 \begin{tabular}{||c c c c c c l c ||}
 \hline
 SNR & $\epsilon_{\rm west}$ & $\epsilon_{\rm east}$ & $\epsilon_{\rm ears}$ &
 $E_{\rm SN,min}$ & $E_{\rm SN,max}$ & Age [\yrs] & References for $E_{\rm SN}$   \\ [0.5ex]
 \hline
 Cassiopeia A & 0.038 & 0.064  & 0.1 & 2 & 3 & 350 & \cite{LamingHwang2003} \\  
& & & & & & &   \cite{Orlandoetal2016}  \\ 
 3C58 &  0.037 & 0.032 & 0.07 & 0.001 & 0.009 & 835 &  \cite{Bocchinoetal2001} \\
Puppis A &  0.009 & 0.010 & 0.02 & 1 & 1 & 1990 & \cite{WinklerPetre2007} \\
S147 &  0.039 & 0.072 & 0.11 & 1 & 3 & 30,000  & \cite{Katsutaetal2012} \\
Vela &  0.005 & 0.004 & 0.01 & 0.12 & 0.14 & 11,400 & \cite{Bocchinoetal1999} \\ 
& & & & & & &  \cite{Sushchetal2011}  \\
G309.2-00.6 &  0.039 & 0.03 & 0.07 & 1 & 1 & 4000 & \cite{Gaensleretal1998} \\
W44 &  0.034 & 0.029 &  0.06 & 0.7 & 0.9 & 20,000 & \cite{Harrusetal1997}  \\
Crab Nebula & $-$ & 0.034 & 0.03 & 0.05 & 0.15 & 963  & \cite{Kitauraetal2006} \\ 
& & & & & & &  \cite{YangChevalier2015}  \\
W49B  &0.1 &0.2  & 0.3 & 0.12 & 1.2 & 1000 & \cite{Micelietal2008} \\
& & & & & & &  \cite{Shimizuetal2013}  \\
G292.0+1.8 &0.035 &0.05  &0.04 & 0.02 & 0.3 & 3000 & \cite{Hughesetal1994} \\
N49B &0.01 &0.01 & 0.02  & 2.7 & 3.1 & 10,000 &\cite{Hughesetal1998} \\
     & 0.24 &$-$ & 0.24 &  & &  &     \\
 \hline
\end{tabular}
\centering
\caption{The ratio $\epsilon_{\rm ear}$, defined in equation (\ref{eq:epsilon1}) as the extra energy in each ear divided by the explosion energy, for 11 CCSNRs; for the western ear ($\epsilon_{\rm west}$) and eastern ear ($\epsilon_{\rm east}$), in the  second and third columns, respectively. For W49B and G292.0+1.8 the second column is for the southern ear and the third column is for the northern ear. The quantity $\epsilon_{\rm ears}$ in the fourth column is the combined values of the two ears. The fifth and sixth columns list the minimum and maximum explosion energy of the SNR (in units of $10^{51} \erg$), as we found in the literature listed in the last column. In the seventh column we list the approximate age of the CCSNR in years, as we found in the following references (same order as in the table). \cite{Ghiotto2015}; \cite{Kothes2016}; \cite{Aschenbach2015}; \cite{Gvaramadze2006}; \cite{Reichleyetal1970} + \cite{SushchHnatyk2014}; \cite{Gaensleretal1998}; \cite{Coxetal1999}; \cite{PolcaroMartocchia2006}; \cite{Lopezetal2009};\cite{Winkleretal2009}; \cite{Parketal2003}. {{{{{ We note that the SNRs S147, W44, and N49B are very old. It is expected that the shaping of these could have been substantially influenced by the ISM. }}}}} 
}
\label{table:calculations}
\end{table}
% TTTTTTTTTTTTTTTTTTTTTTTTTTTTTTTTTTTTTTTTTTTTTTTTTTT

\begin{figure} %[H]
\centering
\includegraphics[width=1\textwidth]{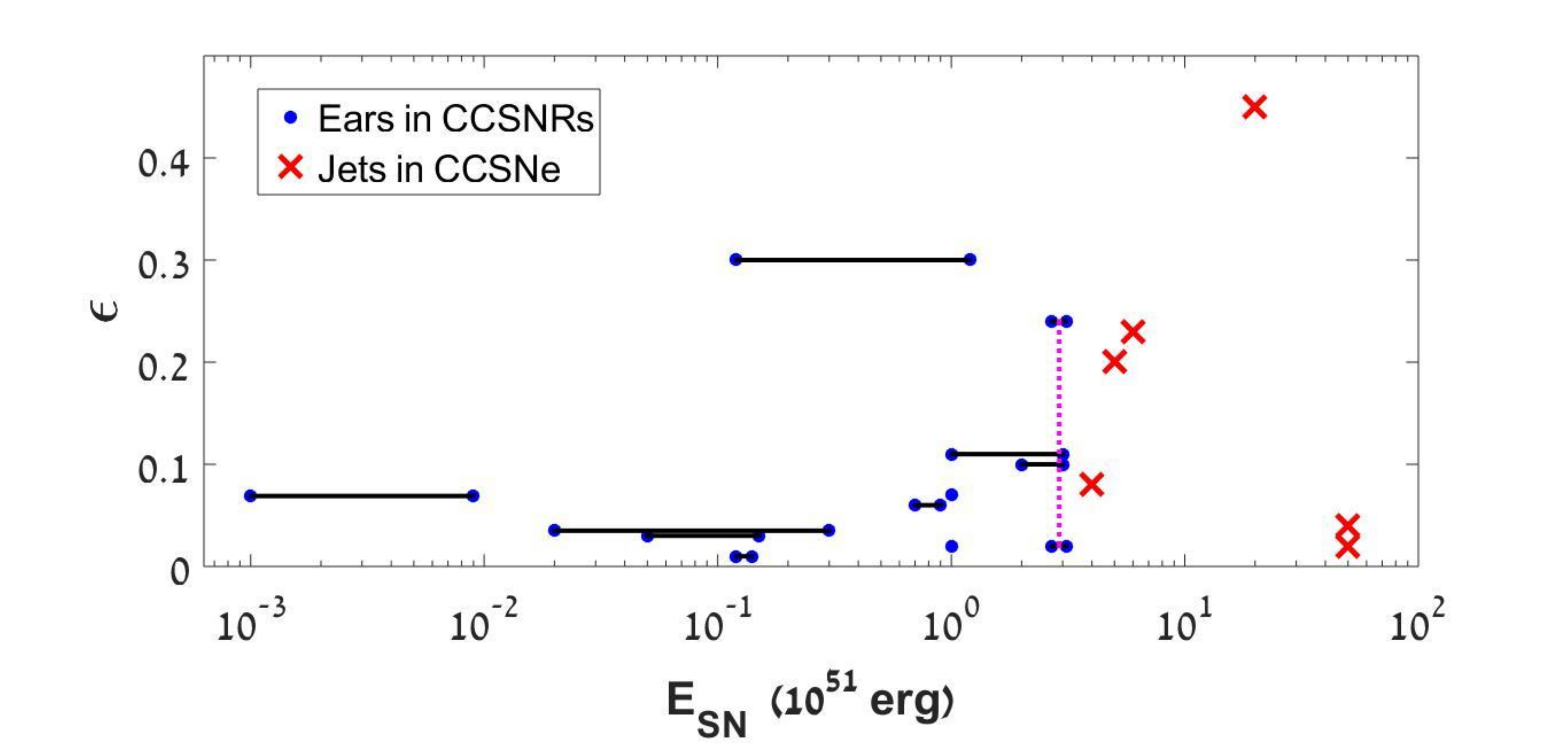}
\caption{Placement of 11 CCSNRs (blue dots) and 6 SLSNe (red crosses) on a diagram of $\epsilon$ (that stands for $\epsilon_{\rm ears}$ or $\epsilon_{\rm SN, jets}$, respectively) versus the explosion energy $E_{\rm SN}$. 
The blue dots represent the ratio of the energy in the two ears combined to that of the SNR shell, $\epsilon_{\rm ears}$, as defined in equation (\ref{eq:epsilon1}) and listed in the fourth column of Table \ref{table:calculations}. The black horizontal lines connect dots for the same SNR of the minimum and maximum values of the explosion energy, 
$E_{\rm {SN,min}}$ and $E_{\rm {SN,max}}$, respectively, that we take from the literature (columns 5-7 in Table \ref{table:calculations}).  The vertical dotted pink line connects the two options for the western ear of N49B. The red crosses represent the ratio of jets to explosion energy 
$\epsilon_{\rm SN,jets} \equiv E_{\rm SN,jets}/E_{\rm SN}$, for the six SLSNe studied by \cite{Piranetal2017}, where the values of $E_{\rm SN,jets}$ and $E_{\rm SN}$ are taken from their Table 1. 
}
\label{figure:EnergyGraph}
\end{figure}

In Fig. \ref{figure:EnergyGraph} we place the 11 CCSNRs we have studied on a plane representing the ratio of the ears energy to SNR shell energy, $\epsilon_{\rm ears}$, as function of the explosion energy. We also place the six CCSNe that were studied by \cite{Piranetal2017}. These are Super Luminous Supernovae (SLSNe) of type Ib/Ic/IIb, and with very energetic explosions. Their progenitor mass was very high.  
\cite{Piranetal2017} obtain the energy in the jets for SLSNe shortly after the explosion, much before the remnant phase. From their estimated energy of the jets, $E_{\rm SN,jets}$ and the explosion energy $E_{\rm SN}$, we calculated the ratio $\epsilon_{\rm SN,jets} \equiv E_{\rm SN,jets}/E_{\rm SN}$ for each of the six objects. We note that there is no correlation in that graph, and the large uncertainties in the explosion energy of some CCSNRs have no influence on our conclusions.  
% FFFFFFFFFFFFFFFFFFFFFFFFFFFFFFFFFFFFFFFFFFFFFFFFFFFFFFFFFF
% FFFFFFFFFFFFFFFFFFFFFFFFFFFFFFFFFFFFFFFFFFFFFFFFFFFFFFFFFF

 The $\epsilon - E_{\rm SN}$ diagram in Fig. \ref{figure:EnergyGraph} (where $\epsilon$ stands for either $\epsilon_{\rm ears}$ or $\epsilon_{\rm SN,jets}$) teaches as the following. 
(1) The ratio of the ears/jets energy to the explosion energy in all SLSNe and CCSNRs is less than unity, $\epsilon <1$. This is compatible with the expectation of the JFM (section \ref{sec:last}). 
(2) When we consider that the energy of the ears is a minimum value for the energy of the jets that inflated the ears, we find that the general range of the ratio $\epsilon$ is the same for CCSNRs (where the ratio is based on ears) and for SLSNe (where the ratio is calculated from the explosion properties). 
(3) The ratio $\epsilon$ does not depend much on the explosion energy over more than three orders of magnitude (hence the uncertainties in the explosion energy of some CCSNRs is of no significance here). 

%==========================================================
\section{SUMMARY}
\label{sec:summary}
%========================================================

We analyzed the morphology of three CCSNRs to obtain some clues on the explosion mechanism of CCSNe. Based on the presence of two opposite protrusions, termed ears, from the SNR main shell (Figs. \ref{G292.0+1.8} and \ref{N49B.eps}) and from the similarities of these three CCSNRs to the morphologies of many PNe (Fig. \ref{RCW103}), we argued that two opposite jets that were launched shortly after the explosion of the core shaped these CCSNRs. We grouped these three CCSNRs together with nine CCSNRs that we studied in previous papers, where we also argued for shaping by jets \citep{GrichenerSoker2017, BearSoker2017}.   

\textit{The new morphologies strengthen our earlier claims that jets play a role in the explosion mechanism of CCSNe (see section \ref{sec:intro}).}

We found the ratio $\epsilon_{\rm ear}$ of the extra kinetic energy of each ear to the explosion energy $E_{\rm SN}$ (equation \ref{eq:epsilon1}), which we took to be the kinetic energy of the SNR ejecta. We listed these values, and the two values combined $\epsilon_{\rm ears}$, for the two CCSNRs with ears studied here in the last rows of Table \ref{table:calculations}. For the CCSNR N49B we listed energies for the small and large western ear. We then placed the two new CCSNRs and the nine CCSNRs we studied in our earlier papers on the $\epsilon_{\rm ears} - E_{\rm SN}$ plane (Fig. \ref{figure:EnergyGraph}).  
We also placed on the $\epsilon_{\rm SN, jets} - E_{\rm SN}$ plane the six SLSNe studied by \cite{Piranetal2017}, where in that case instead of the ears energy we took the energy of the two jets to calculate the value of $\epsilon_{\rm SN, jets}$.  

From Fig. \ref{figure:EnergyGraph} we learn that the relative energies of the jets that are detected in energetic CCSNe (SLSNe) and those that are inferred to shape ears in CCSNRs share the same range, and that this relative jets' energy is less than unity, $\epsilon_{\rm ears} < 0.5$ and $\epsilon_{\rm jets} < 0.5$ in the figure.   
 As we explained in section \ref{sec:last}, in the jet feedback explosion mechanism of CCSNe only the jets of the last launching episode (or last two episodes) have their imprints on the outskirts of the expanding gas. The energy in these jets is from about a percent to tens of percents of the explosion energy. 
 
\textit{We conclude that the finding that jets that leave their imprints on the outer ejecta of regular CCSNe and of SLSNe carry only a small fraction of the explosion energy, is compatible with the jet feedback explosion mechanism of massive stars. }
 
 We do point out that although in this paper we discuss the imprints of jets on the outer parts of the ejecta, the opposite can also take place. In the CCSNR RCW~103, for example (Fig. \ref{RCW103}), the jets influenced the entire SNR, along the faint region from the center out in the two opposite directions marked by the yellow arrows. In cases when the last jets strongly jitter and/or are of very short duration, they will leave imprints in the inner regions rather than in the outer regions. 

The usage of CCSNRs can reveal more information on the explosion mechanism. In a future work we plan to address the relative motion of the NS and the SNR within the frame of the JFM. 
For example, in a recent paper \cite{HollandAshfordetal2017} compare the morphology of CCSNRs with the motion of the NS at their center. We note that four out of the six SNRs that they show in their figure 4 have been studied by us for jets, here and in previous papers. From the direction of the jets that we deduce and the NS motion given by \cite{HollandAshfordetal2017} in each of these four CCSNRs, we have tentative indications that the natal kick of the NS is more or less perpendicular to the axis of the last jets. 

We end the present study by reiterating our call for a paradigm shift from neutrino-driven explosion to a jet-driven explosion (e.g., \citealt{Papishetal2015b, Soker2017a}). 
 
 We thank an anonymous referee for detail and helpful comments. This research was supported by the Israel Science Foundation and a grant from the Asher Space Research Institute at the Technion. A.G. was supported by The Rothschild Scholars Program- Technion Program for Excellence.

\end{document}